\documentclass[12pt]{article}

\usepackage{amssymb}
\usepackage{amsmath}
\usepackage{amscd}
\usepackage{latexsym}
\usepackage{graphicx}

\usepackage{cite}

\newcommand{\be}{\begin{equation}}
\newcommand{\ee}{\end{equation}}
\newcommand{\Dlt}{\Delta}
\newcommand{\dlt}{\delta}
\newcommand{\prt}{\partial}
\newcommand{\br}{{\bf r}}

\newcommand{\vp}{\varphi}

\newcommand{\al}{\alpha}
\newcommand{\ra}{\rightarrow}

\newcommand{\om}{\omega}

\newcommand{\lbd}{\lambda}

\begin{document}

\begin{center}

{\Large{\bf Characteristic quantities for nonequilibrium Bose systems} \\ [5mm]

V.I. Yukalov$^{1,2}$, A.N. Novikov$^3$, E.P. Yukalova$^{4}$ and V.S. Bagnato$^2$} \\ [3mm]

{\it
$^1$Bogolubov Laboratory of Theoretical Physics, \\
Joint Institute for Nuclear Research, Dubna 141980, Russia \\ [2mm]

$^2$Instituto de Fisica de S\~ao Carlos, Universidade de S\~ao Paulo, \\
CP 369,  S\~ao Carlos 13560-970, S\~ao Paulo, Brazil   \\ [2mm]
 
$^3$Colegio de Ciencias e Ingenierias, Universidad San Francisco de Quito, \\
Quito, Ecuador \\ [2mm]

$^4$Laboratory of Information Technologies, \\
Joint Institute for Nuclear Research, Dubna 141980, Russia } \\ [5mm]

{\bf E-mail}: yukalov@theor.jinr.ru

\end{center}

\vskip 5cm

\begin{abstract}
The paper discusses what characteristic quantities could quantify nonequilibrium 
states of Bose systems. Among such quantities, the following are considered: 
effective temperature, Fresnel number, and Mach number. The suggested classification 
of nonequilibrium states is illustrated by studying a Bose-Einstein condensate in 
a shaken trap, where it is possible to distinguish eight different nonequilibrium 
states: weak nonequilibrium, vortex germs, vortex rings, vortex lines, deformed 
vortices, vortex turbulence, grain turbulence, and wave turbulence. Nonequilibrium 
states are created experimentally and modeled by solving the nonlinear Schr\"odinger 
equation. 
\end{abstract}

\newpage

\section{Generation of nonequilibrium condensates}

Weakly interacting atoms at asymptotically low temperature are almost completely 
Bose condensed, forming a coherent system. Such a system can be described in the 
quasiclassical approximation resulting in the equation for the coherent field 
corresponding to the  nonlinear Schr\"{o}dinger (NLS) equation
\be
\label{1}
 i\; \frac{\prt}{\prt t}\; \psi(\br,t) = \hat H[\;\psi\;] \; \psi(\br,t) \; ,
\ee
in which the nonlinear Hamiltonian is
\be
\label{2}
 \hat H[\;\psi\;] = -\; \frac{\nabla^2}{2m} +U(\br,t) + N
\int \Phi(\br-\br') \; |\; \psi(\br',t) \;|^2\; d\br' \; .
\ee
Here the Planck constant is set to one, $U({\bf r},t)$ is an external potential, and 
$\Phi({\bf r})$ is the interaction potential of atoms. This equation was advanced by 
Bogolubov \cite{Bogolubov_1} in 1949 in his widely known book Lectures on Quantum 
Statistics (see also \cite{Bogolubov_2,Bogolubov_3,Bogolubov_4}). 

For a dilute gas, the interaction potential is modeled by the local form
\be
\label{3}
  \Phi(\br) = \Phi_0\dlt(\br) \qquad 
\left( \Phi_0 \equiv 4\pi\; \frac{a_s}{m}\right) \; ,
\ee
where $a_s$ is scattering length. Then the nonlinear Hamiltonian becomes
\be
\label{4}
 \hat H[\;\psi\;] = -\; \frac{\nabla^2}{2m} +U(\br,t) + 
N \Phi_0 |\; \psi(\br,t) \;|^2  \; .
\ee

The solution to the NLS equation is also called the condensate wave function. In the 
present case, it is normalized to one, 
\be
\label{5}
 \int |\; \psi(\br,t) \;|^2 \; d\br = 1 \; .
\ee
The stationary solution to the NLS equation corresponds to the situation where the 
external potential does not depend on time, or simply is absent, and the initial 
condition is also stationary. Then the substitution 
$$
  \psi(\br,t) \ra \vp(\br) e^{-iE t} 
$$ 
yields the stationary NLS equation
$$
\hat H[\;\vp \;] \vp(\br) = E \vp(\br) \; .
$$
The minimal $E$ corresponds to the ground-state energy of the Bose-condensed system. 

Nonstationary solutions arise when either the initial condition is nonstationary 
or when the external potential depends on time. The creation of a nonequilibrium 
condensate can be done by applying external fields modulating either the trap 
potential or the scattering length. This way allows us to create not merely weakly 
nonequilibrium condensates, but strongly nonequilibrium condensates containing 
coherent topological modes \cite{Yukalov_5,Yukalov_6,Yukalov_7}. 

The creation of strongly nonequilibrium condensates has been studied both 
experimentally and in numerical modeling. In the experiments of the group from the 
University of S\~ao Paulo, trapped atoms of $^{87}$Rb, with the scattering length 
$a_s=0.557 \times 10^{-6}$ cm were employed \cite{Shiozaki_8,Seman_9}. A cylindrical 
trap with the radial frequency $\om_r=2\pi\times 210$ Hz and longitudinal frequency 
$\om_z=2\pi\times 23$ Hz was used. The trap is elongated, with the aspect ratio 
$\al\equiv\om_z/\om_r=0.11$. The number of condensed atoms in the trap is 
$N\approx 1.5 \times 10^5$. The temporal modulation was accomplished through the 
trapping potential of the form
$$
U(\br,t) = \frac{m}{2}\; \om_r^2 ( x\cos\Theta_2 + z \sin\Theta_2 - 
\dlt_3 B_t )^2 \; +
$$
\be
\label{6}
+ \; \frac{m}{2}\; \om_r^2 ( y\cos\Theta_1 - z \sin\Theta_1 - \dlt_2 B_t )^2 \; + 
\; \frac{m}{2}\; \om_z^2 ( z\cos\Theta_1 -x \sin\Theta_2 + y \sin\Theta_1 - 
\dlt_1 B_t )^2 \;  ,
\ee
where
$$
\Theta_i = A_i ( 1 -\cos\om t) \; , \qquad B_t = 1 -\cos\om t \;   .
$$
The modulation frequency is $\omega = 2\pi \times 200$ Hz and the modulation 
amplitude is about $0.2\om_r$.     

All parameters of the setup were taken the same in the numerical simulations and 
in experiments. The results of numerical simulations are close to the experimentally 
observed. Here we concentrate on numerical simulations, since they allow us to clearly 
illustrate the sequence of created nonequilibrium states and to calculate the related
characteristic quantities.

\section{Numerical simulation} 

The numerical solution of Eq. (\ref{1}) is accomplished by using the methods thoroughly 
described in \cite{Novikov_10}). Here we demonstrate the resulting nonequilibrium states 
in more details than in the previous works 
\cite{Yukalov_11,Yukalov_12,Yukalov_13,Yukalov_14}. The following nonequilibrium 
states have been observed. At the first stage, lasting during the time interval 
$0 < t < 5$ ms, when the condensate is yet weakly disturbed, there appear only 
density fluctuations, without forming complicated structures.

For longer modulation of the trapping potential, in the time interval $5<t<10$ ms,
there arise the germs of vortex rings shown in Fig. 1. If the pumping is stopped 
at this stage, the germs survive during the time around $0.2$ s, which shows that 
they are metastable topological objects.    

In the interval $10<t<15$ ms, well defined rings appear in pairs, with opposite 
circulations, so that their total circulation is zero. The circulation number for 
each ring is $\pm 1$. The corresponding picture is shown in Fig. 2. The ring 
lifetime, after switching off the pumping, is about $0.1$ s.

For $15<t<17$ ms, vortex lines arise (left Fig 3), with $\pm 1$ circulation number. 
The total circulation in the system is zero. A vortex lifetime is $0.2$ s. After 
$17$ ms, vortices start deforming, as in the right Fig. 3, and become strongly 
deformed in the interval $19<t<25$ ms, as is shown in Fig. 4.   

In the time interval $25<t<27$ ms, the stage of vortex turbulence begins to develop, 
with strongly deformed and entangled vortices (Fig. 5). The dots in the figure are 
assumed to be connected with their nearest neighbours, which is rather difficult to 
draw, since the vortices are so strongly entangled and deformed. The turbulent regime 
becomes well developed after $27$ ms. The well developed turbulence exists approximately 
in the interval $27<t<30$ ms (Fig. 6), after which vortices start decaying (Fig. 7). 
The decaying turbulence occurs in the interval $30<t<40$ ms. Thus in total, the stage 
of quantum turbulence exists in the time interval $25<t<40$ ms. The state of quantum 
turbulence possesses the standard properties, typical of Vinen turbulence 
\cite{Vinen_15,Vinen_16,Tsubota_17,Nemorovskii_18,Tsatsos_18}, such as the occurrence 
of a random vortex tangle and an anisotropic self-similar expansion preserving the aspect 
ratio of the cloud during its expansion. The number of vortices as a function of time, 
characterizing the beginning, development, and decay of turbulence, is shown in Fig. 8. 
 
After about $45$ ms, the number of vortices quickly diminishes almost to zero and the 
granular, or droplet, state develops. In this state, the system is formed by dense 
droplets (grains) that are randomly distributed in space and surrounded by a rarified 
gas. The density inside a droplet is $100$ times larger than in the surrounding. The 
typical radius of a droplet is about $1.5 \times 10^{-5}$ cm, which is close to the 
coherence length $\xi$. The phase inside a droplet is constant, while in the 
surrounding gas the phase is random, which implies that each droplet is a coherent 
object. The lifetime of a droplet, after switching off pumping, is much longer than 
the local equilibrium time,
$$
 t_{life} \sim 10^{-2}\; {\rm s} \gg t_{loc} \sim 10^{-3} \; {\rm s} \; ,
$$
which means that a droplet is a metastable formation. Generally, the droplet state 
has to satisfy the following properties in order to be classified as such. (1) the 
typical size of each droplet is of the order of the healing length; (2) the phase 
inside a droplet is constant, which defines a droplet as a coherent object; (3) the 
phase in the space around a droplet is random; (4) the lifetime of a droplet has to 
be much longer than the local equilibrium time; (5) the density inside a coherent 
droplet should be much larger than that of its incoherent surrounding. Under the 
considered pumping, the droplet state lasts approximately in the interval 
$40<t<150$ ms. 

By the end of the droplet regime, after $150$ ms, the number of droplets falls and 
the regime of wave turbulence develops, where the system consists of small-amplitude 
waves of the density only three times larger than in the surrounding. The typical size 
of each wave is around $10^{-4}$ cm. The phase is everywhere random, so that there is 
no coherence either inside each wave or outside it. The density snapshots comparing 
the state of an equilibrium condensate with the droplet state and the state of wave 
turbulence is presented in Fig. 9.

\section{Characteristic quantities}

The appearance of this or that state is clearly connected with the amount of energy 
pumped into the system. The amount of energy transferred to the system depends on 
the amplitude of perturbation and on the duration of the perturbation time. It is, 
therefore, more convenient to distinguish different regimes not merely by the pumping 
time, but rather by the amount of the energy pumped into the system. The alternating 
external field mainly increases the kinetic energy of the system, as is seen from 
Fig. 10. Hence it is logical to introduce such characteristics that take into account 
the change in the system kinetic energy \cite{Yukalov_19}. The change of kinetic energy 
can be connected with the effective temperature
\be
\label{7}
 T_{eff} = \frac{2}{3}\; \left( E_{kin} - E_{kin}^{eq} \right) = 
\frac{2}{3}\;\Dlt E_{kin}   
\ee
defined through the difference of the current-state kinetic energy and its value at 
equilibrium. 
  
The states of such nonequilibrium systems as lasers are usually characterized by 
Fresnel number, which for a system of cylindrical shape of radius $R$ and length $L$ 
reads
\be
\label{8}
 F = \frac{A_{cross}}{\lbd L}  = \frac{\pi R^2}{\lbd L}  \;  .
\ee
Here, under the wavelength, it is possible to assume the thermal wavelength
\be
\label{9}
 \lbd = \sqrt{ \frac{2\pi}{m T_{eff} } } \;  ,
\ee
which results in the effective Fresnel number
\be
\label{10}
 F = \sqrt{ \frac{\pi\al\Dlt E_{kin}}{12\hbar \om_r } } \;  ,
\ee
with the aspect ratio
$$
 \al \equiv \frac{\om_z}{\om_r} = \frac{4R^2}{L^2} \; .
$$
Finally, it is possible to introduce the effective Mach number
\be
\label{11}
 M = \frac{v}{c} = \sqrt{ \frac{2\Dlt E_{kin}}{mc^2} } \;  ,
\ee
where $c$ is sound velocity. The classification of nonequilibrium states by means 
of the introduced characteristic quantities is illustrated in Fig. 11. Note that 
the developed regime of wave turbulence, where coherence completely disappears, 
corresponds to the effective temperature $T_{eff}=23.5\om_r$, which practically 
coincides with the critical temperature $T_c=23.8\om_r$ for $^{87}$Rb in the 
studied setup.

\section{Resonant generation}  

The frequency of the alternating field modulating the trap potential in the above 
setup was not connected to any transition frequency of the Bose system in the trap. 
If, however, the modulation frequency is specially tuned to the transition frequency 
between the ground state and an excited state representing a coherent topological 
mode, as described in Refs. \cite{Yukalov_5,Yukalov_6,Yukalov_7}, the beginning of 
the nonequilibrium process, for times $t\ll 100$ ms, can be different. This is 
illustrated in Figs. 12 and 13 by the temporal behaviour of the Fresnel number for 
different aspect ratios and different modulation amplitudes, with other parameters 
fixed as in all calculations above. In these figures, the parameter $b$ defines the 
ratio of the modulation amplitude to the strength of interactions between the atoms. 
The effective Fresnel number oscillates with time. The amplitude of the oscillations 
and their period increase with growing modulation amplitude $b$. In contrast, when 
$b$ is fixed and the aspect ratio increases, only the oscillation amplitude rises, 
while the oscillation period does not vary.  

The resonant generation is possible only at the beginning of the pumping process, 
till times of order $10$ ms. For much larger times, the effect of power broadening 
comes into play, and the whole process of creation of nonequilibrium states continues 
approximately as under nonresonant pumping.        

In conclusion, we demonstrated that strongly nonequilibrium states of Bose-Einstein 
condensates, containing topological modes, can be created by subjecting the system 
of trapped atoms to external alternating fields. The generated nonequilibrium states 
are conveniently characterized by effective temperature, Fresnel number, and Mach 
number.

\newpage

\begin{figure}[ht]
\centerline{
\includegraphics[width=10cm]{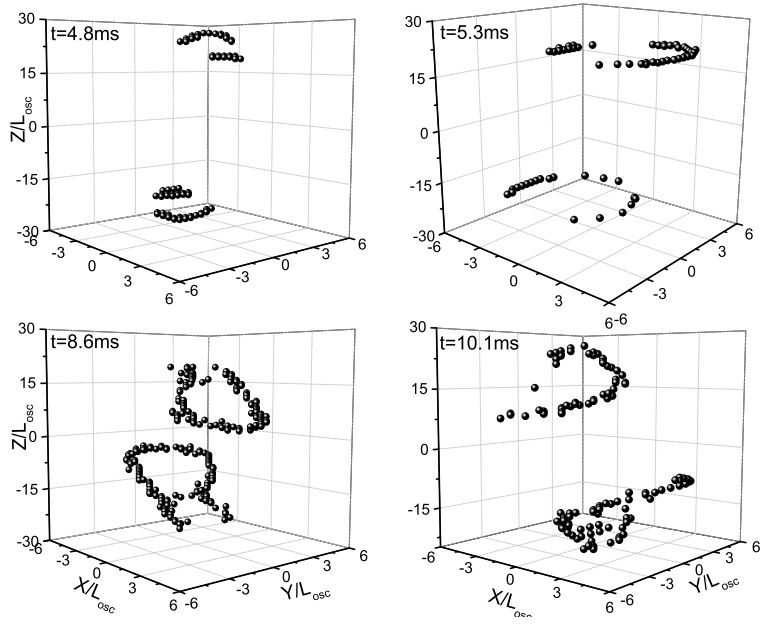} }
\caption{Typical vortex germs.
}
\label{fig:Fig.1}
\end{figure}

\vskip 3cm

\begin{figure}[ht]
\centerline{
\includegraphics[width=10cm]{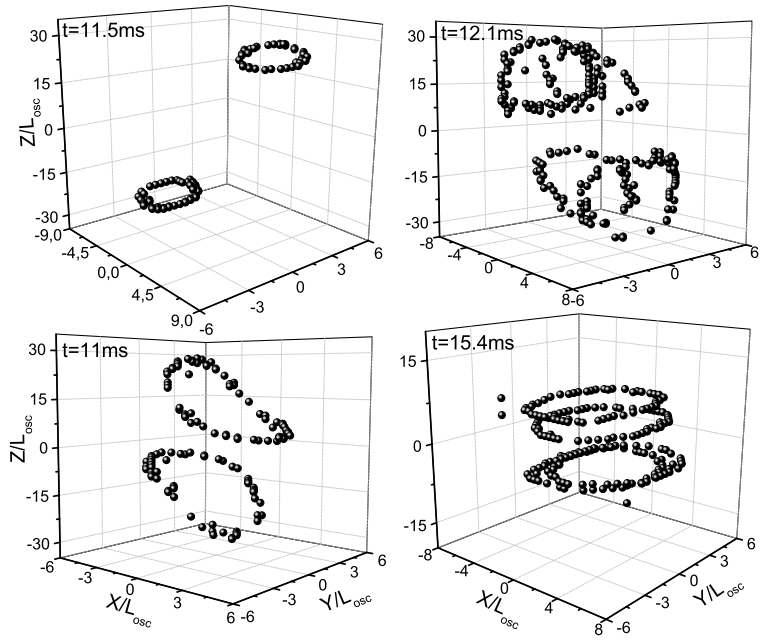} }
\caption{Typical vortex rings.
}
\label{fig:Fig.2}
\end{figure}

\newpage

\begin{figure}[ht]
\centerline{
\includegraphics[width=12cm]{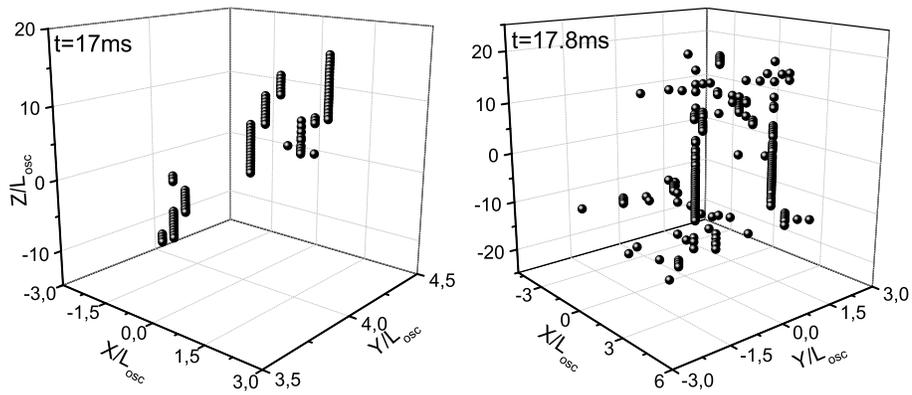} }
\caption{Almost straight vortex lines (left figure) and slightly deformed vortices 
(right figure).
}
\label{fig:Fig.3}
\end{figure}

\vskip 3cm

\begin{figure}[ht]
\centerline{
\includegraphics[width=12cm]{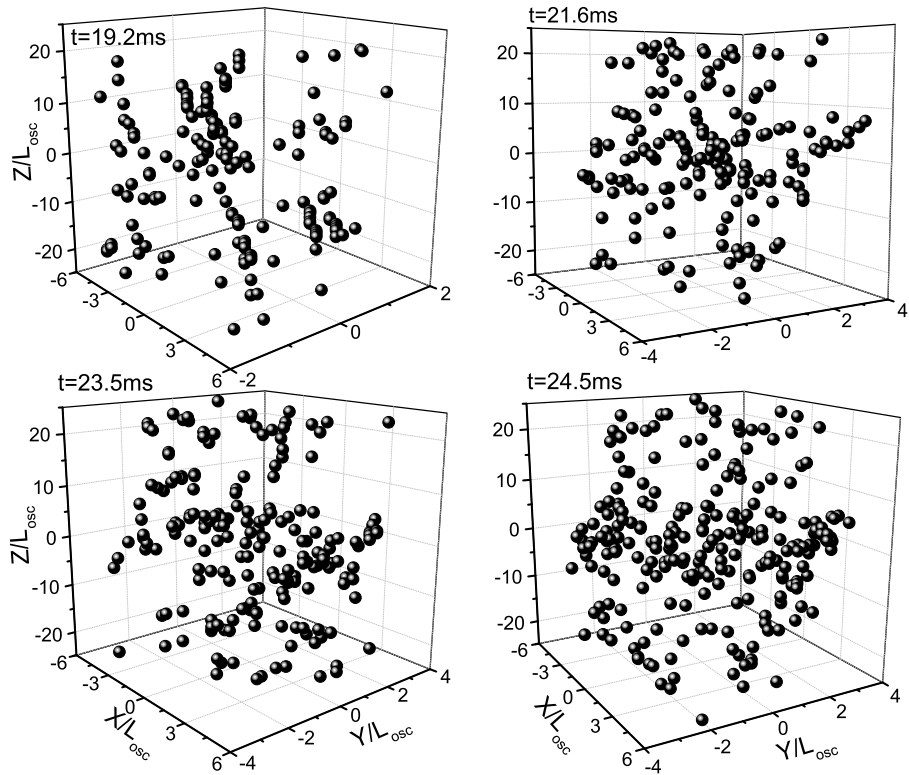} }
\caption{Strongly deformed vortices.
}
\label{fig:Fig.4}
\end{figure}

\newpage

\begin{figure}[ht]
\centerline{
\includegraphics[width=14cm]{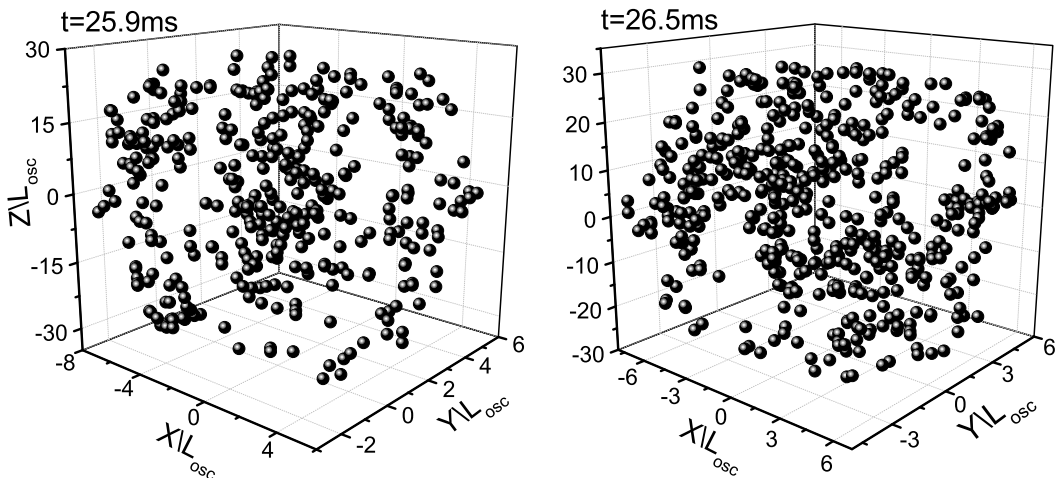} }
\caption{Initial stage of vortex turbulence.
}
\label{fig:Fig.5}
\end{figure}

\vskip 3cm

\begin{figure}[ht]
\centerline{
\includegraphics[width=14cm]{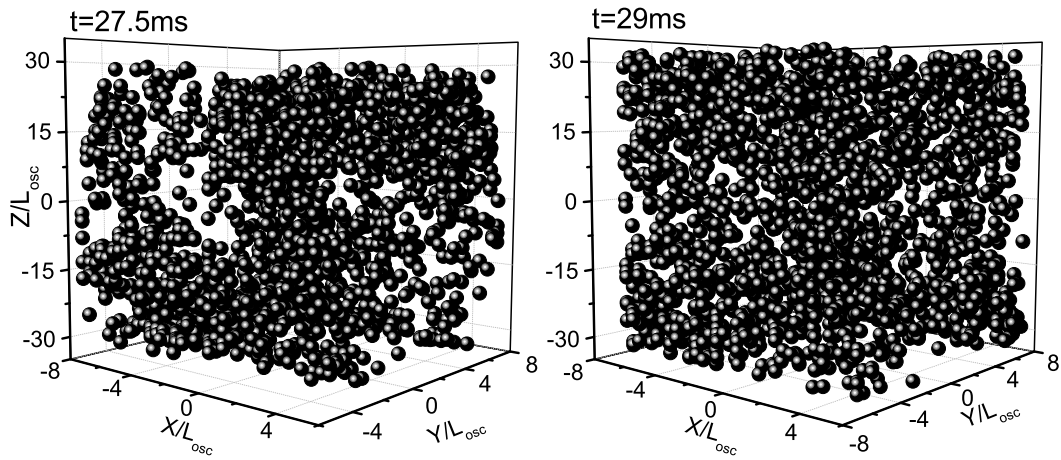} }
\caption{Developed vortex turbulence.
}
\label{fig:Fig.6}
\end{figure}

\newpage

\begin{figure}[ht]
\centerline{
\includegraphics[width=10cm]{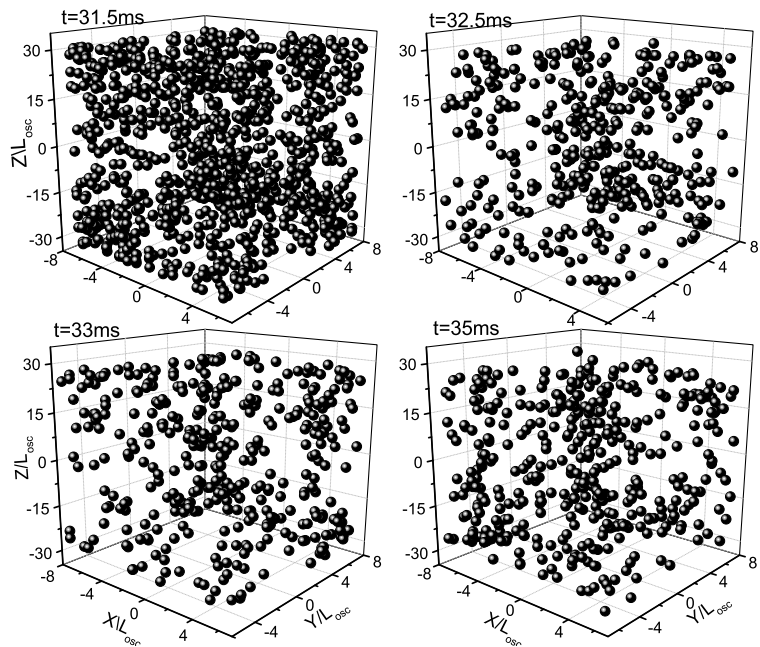} }
\caption{Decaying vortex turbulence.
}
\label{fig:Fig.7}
\end{figure}

\vskip 3cm

\begin{figure}[ht]
\centerline{
\includegraphics[width=8cm]{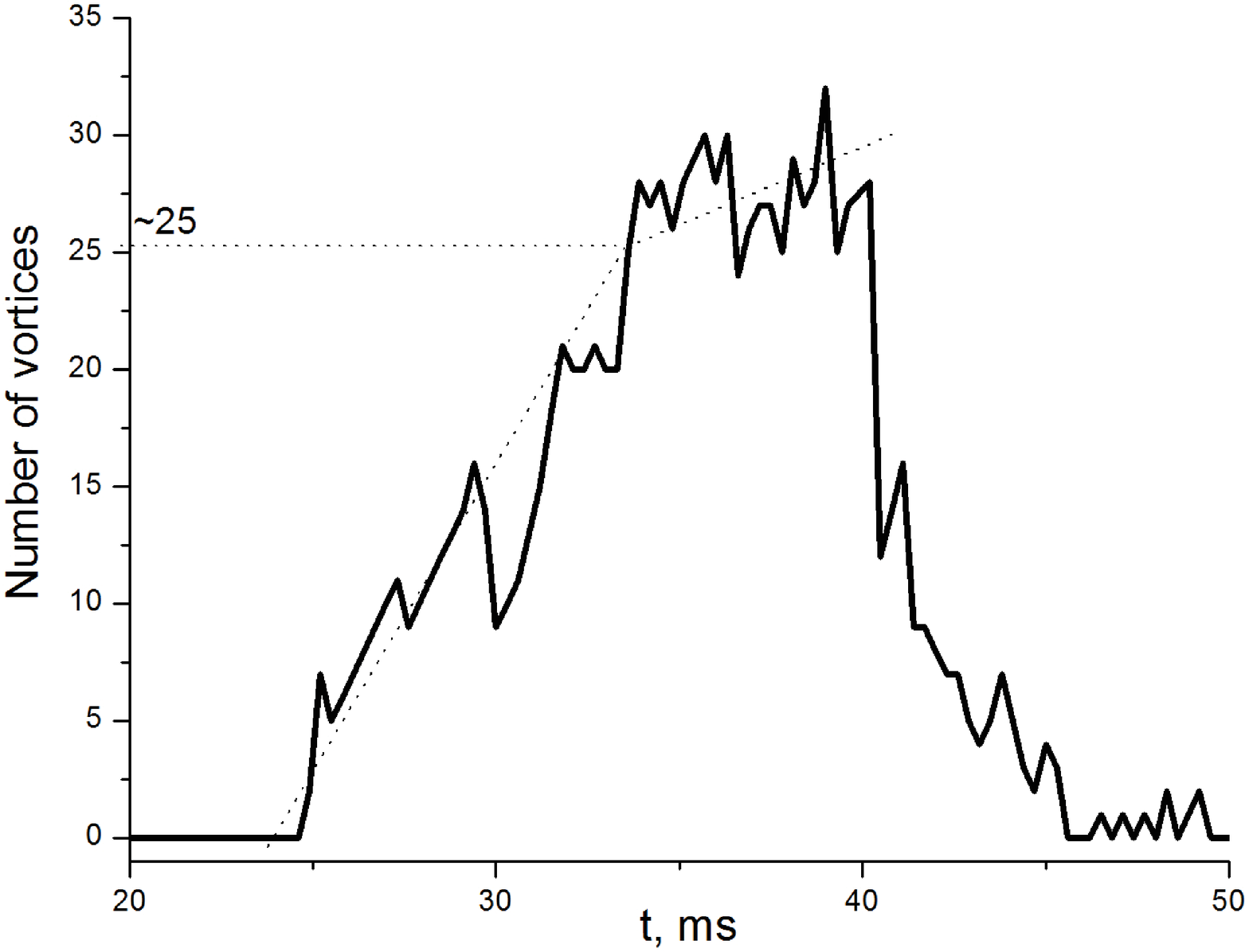} }
\caption{Number of vortices as a function of time.
}
\label{fig:Fig.8}
\end{figure}

\newpage

\begin{figure}[ht]
\centerline{
\includegraphics[width=14cm]{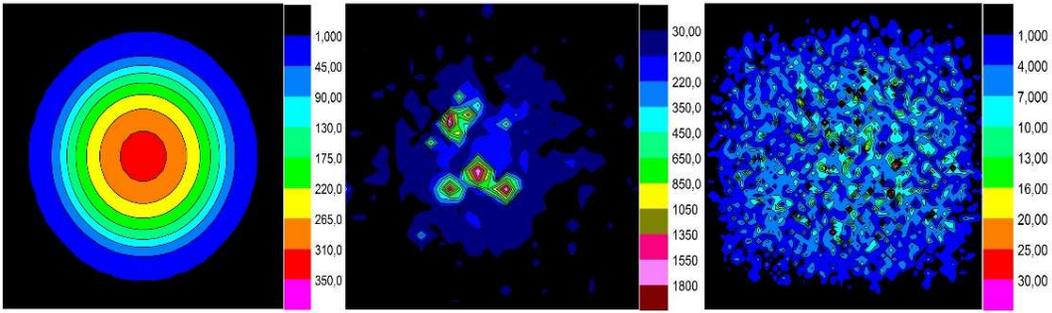} }
\caption{Density snapshot of equilibrium condensate (left plot), droplet state 
(middle plot), and wave turbulence (right plot).
}
\label{fig:Fig.9}
\end{figure}

\vskip 3cm

\begin{figure}[ht]
\centerline{
\includegraphics[width=8cm]{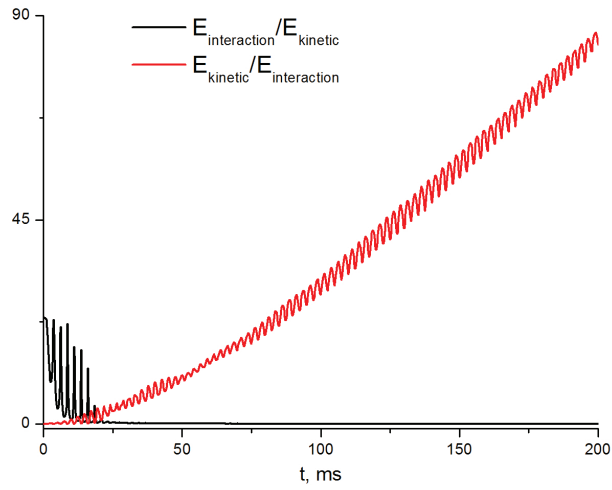} }
\caption{Comparison of kinetic and potential energies in the condensate of 
trapped atoms.
}
\label{fig:Fig.10}
\end{figure}

\newpage

\begin{figure}[ht]
\centerline{
\includegraphics[width=15cm]{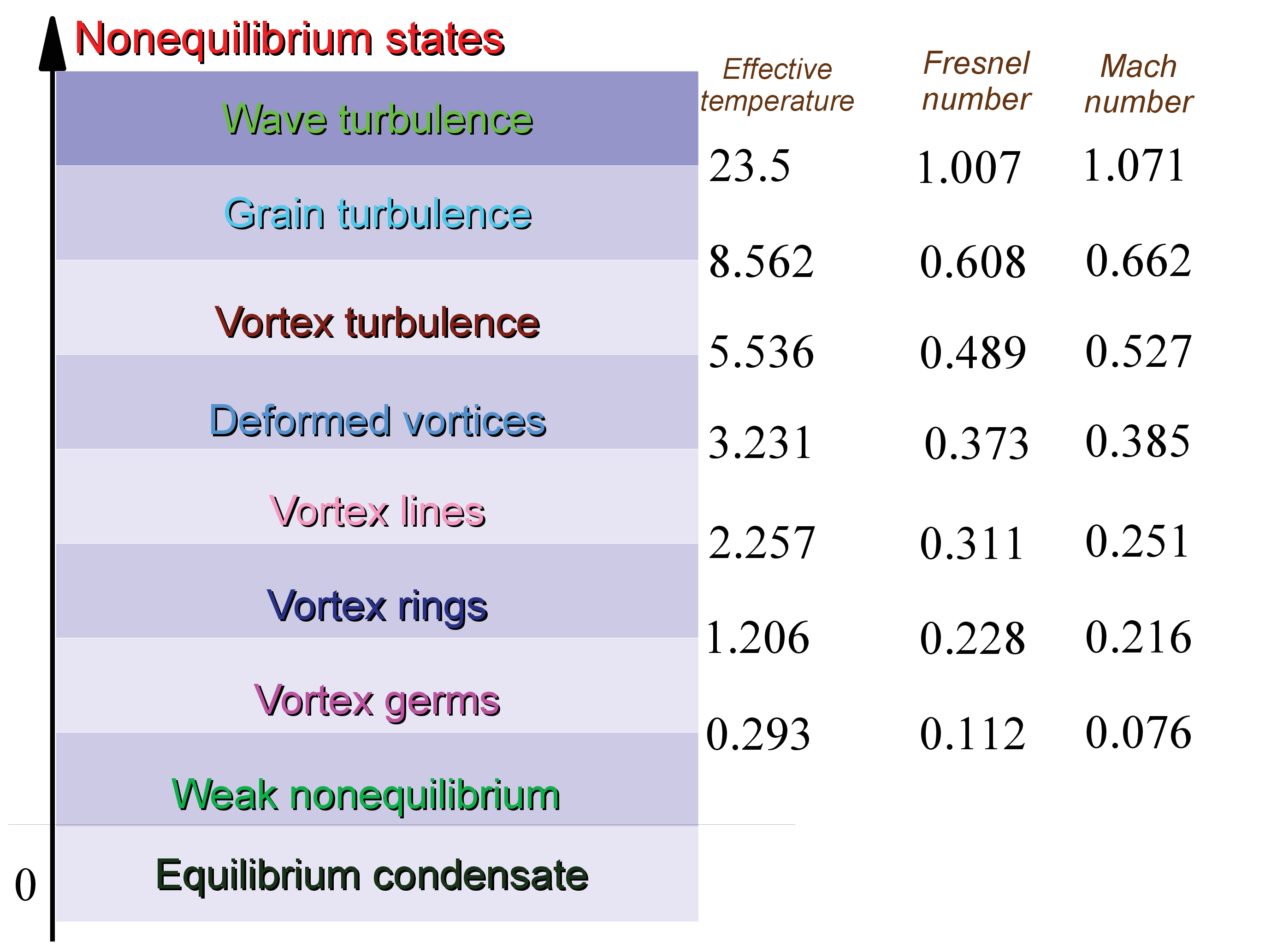} }
\caption{Classification of nonequilibrium states of trapped Bose-condensed atoms.
}
\label{fig:Fig.11}
\end{figure}

\newpage

\begin{figure}[ht]
\centerline{
\includegraphics[width=10cm,height=7cm]{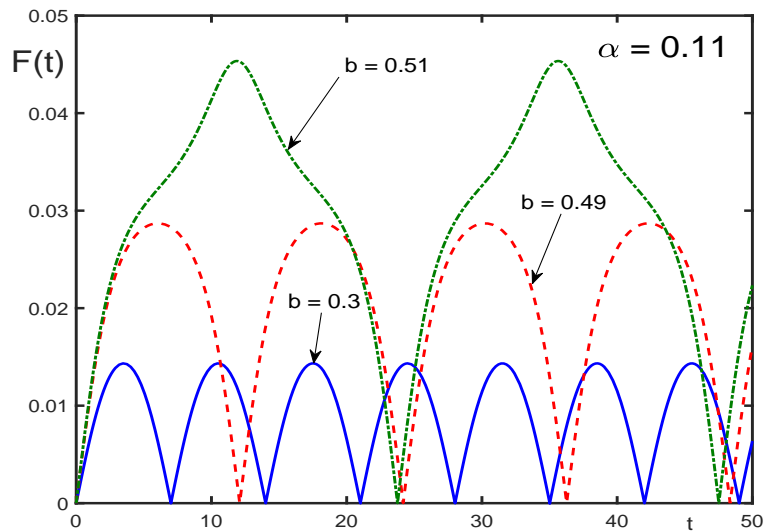} }
\caption{Temporal variation of Fresnel number for a fixed aspect ratio $\al$ and 
different modulation amplitudes, labeled by the parameter $b$.
}
\label{fig:Fig.12}
\end{figure}

\vskip 3cm

\begin{figure}[ht]
\centerline{
\includegraphics[width=10cm,height=7cm]{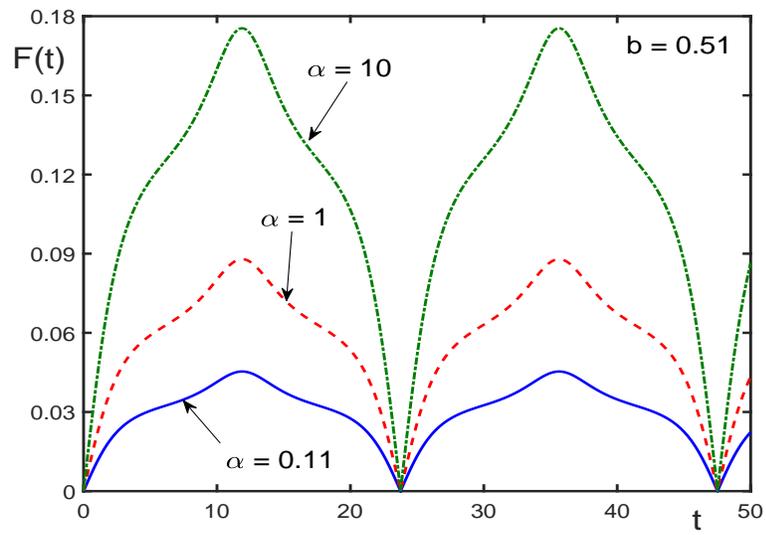} }
\caption{Temporal variation of Fresnel number for a fixed modulation amplitude and 
different values of aspect ratio.
}
\label{fig:Fig.13}
\end{figure}

\end{document}